\documentclass[copyright,creativecommons,noderivs,noncommercial]{eptcs}

\usepackage{tikz}
\usetikzlibrary{snakes,arrows}

\usepackage{url}
\usepackage{fancyvrb, courier} 

\DefineVerbatimEnvironment{code}{Verbatim}{frame=none, framesep=10mm, fontsize=\small, commandchars=\\\{\}}
\usepackage{graphicx}
\usepackage{caption}
\usepackage{subcaption}
\usepackage{listings,xcolor}
\newcommand{\var}[1]{{\textit{#1}}}
\definecolor{codebg}{gray}{0.93}

\lstnewenvironment
  {ttmcodel}[1][]
  {\lstset{
     language=TTM,
     columns=fullflexible,
     mathescape=true,
     #1,
	 basicstyle=\footnotesize,}}
  {}

\usepackage{multirow} 
\usepackage{bsymb}
\usepackage{amsmath, amssymb}
\def\spot{\,\bullet\,}
\usepackage{wrapfig}

\let\OLDthebibliography\thebibliography
\renewcommand\thebibliography[1]{
  \OLDthebibliography{#1}
  \setlength{\parskip}{0pt}
  \setlength{\itemsep}{0pt plus 0.3ex}
}

\makeatletter
 \newcommand\npar{\@startsection{subsection}{2}{\z@}{-2\p@ \@plus -4\p@ \@minus -4\p@}{-0.5em \@plus -0.22em \@minus -0.1em}{\normalfont\large\bfseries}}
 \makeatother

\usepackage{hyperref}
\hypersetup{%
  colorlinks=false,
  linkbordercolor=red,
  pdfborderstyle={/S/U/W 1}
}

\begin{document}
	
\title{Using Indexed and Synchronous Events to \\ Model and Validate Cyber-Physical Systems}
\author{Chen-Wei Wang, Jonathan S. Ostroff, and Simon Hudon
\institute{Department of Electrical Engineering and Computer Science, \\ York University, Canada}
\email{\{jackie, jonathan, simon\}@cse.yorku.ca}} 

\def\authorrunning{Wang, Ostroff, and Hudon}
\def\titlerunning{Indexed Events \& Synchronous Events in TTM}

\maketitle

\begin{abstract}
Timed Transition Models (TTMs) are event-based descriptions for modelling, specifying, and verifying discrete real-time systems. An event can be spontaneous, fair, or timed with specified bounds. TTMs have a textual syntax, an operational semantics, and an automated tool supporting linear-time temporal logic. We extend TTMs and its tool with two novel modelling features for writing high-level specifications: indexed events and synchronous events. Indexed events allow for concise description of behaviour common to a set of actors.  The indexing construct allows us to select a specific actor and to specify a temporal property for that actor. We use indexed events to validate the requirements of a train control system. Synchronous events allow developers to decompose simultaneous state updates into actions of separate events. To specify the intended data flow among synchronized actions, we use primed variables to reference the post-state (i.e., one resulted from taking the synchronized actions). The TTM tool automatically infers the data flow from synchronous events, and reports errors on inconsistencies due to circular data flow. We use synchronous events to validate part of the requirements of a nuclear shutdown system. In both case studies, we show how the new notation facilitates the formal validation of system requirements, and use the TTM tool to verify safety, liveness, and real-time properties.
\end{abstract}



\section{Introduction}

Cyber-physical systems integrate computational systems (the ``controller'') with physical processes (the ``plant''). Such systems are found in areas as diverse as aerospace, automotive, energy, healthcare, manufacturing, transportation, and consumer appliances. A main challenge in developing cyber-physical systems is modelling the joint dynamics of computer controllers and the plant~\cite{Derler2012}. 

Timed Transition Models (TTMs) are event-based descriptions for modelling, specifying, and verifying discrete real-time systems. A system is composed of module instances. Each module declares an interface and a list of events. An event can be spontaneous, fair, or timed (i.e., with lower and upper time bounds). In \cite{Ostroff2014}, we provided TTMs with a textual syntax, an operational semantics, and an automated tool, including an editor with type checking, a graphical simulator, and a verifier for linear-time temporal logic. So far, TTMs were used to verify that a variety of implementations satisfy their specifications. 

In this paper, we extend the TTM notation, semantics, and tool for two novel modelling features: indexed events and synchronous events. These constructs are suitable for writing high-level specification, and can thus facilitate the validation of system requirements. 

\textit{Indexed events} allow for concise description of behaviour common to a (possibly unspecified) set of actors. The indexing construct allows us to select a specific actor (such as a train) and specify a temporal property for that actor. For example, let $loc$ be an array of train locations (a train can be on either the entrance block, a platform, an exit block, or outside the station). An event \var{move\_out} can be indexed with a set \var{TRAIN} of trains, which results in an indexed event \var{move\_out(t: fair TRAIN)} describing the action of a train $t$ moving out of a platform and into the exit block. As a result, the event index $t$ can be used to specify the liveness property that every train $t$ waiting at one of the platforms (denoted by the set \emph{PLF}) eventually moves out, and into the exit block: $\Box(\var{loc}[\var{t}] \in \var{PLF} ~\Rightarrow~ \Diamond \var{move\_out}(\var{t}))$. Without the index \var{t}, we can only state a weaker property that some train eventually leaves the station (unless we introduce auxiliary variables or events). 
 
\textit{Synchronous events} allow developers to decompose simultaneous state updates into actions of separate events. However, without a mechanism to reference the post-state values of monitored variables, we cannot properly model the joint actions of the environment and controller. For example, the synchronized action $\var{m} := \var{exp} ~||~ \var{c} := \var{f}~(\var{m})$ specifies that the new (or next-state) value of controlled variable \var{c} is computed on the basis of the old (or pre-state) value of monitored variable \var{m} (i.e., \var{exp}). To resolve this, we use primed variables on the RHS of assignments in event actions to denote post-state values. For example, the synchronized action $\var{m} := \var{exp} ~||~ \var{c} := \var{f}~(\var{m'})$ specifies that the post-state value of \var{c} is now a function on the post-value value of \var{m}. Synchronous events, together with primed variables, are suitable for describing high-level specifications used in shutdown systems of nuclear reactors~\cite{Wassyng2006}. In such systems, the next-state value of the system controlled variables are expressed in terms of the current-state and next-state values of the monitored variables of nuclear reactors. This allows for a simplified description of the requirements that will later be refined to code.

\smallskip

\noindent\textit{Contributions}. To support indexed and synchronous events for validating requirements, we extend the semantics of TTM (Sec.~\ref{semantics}), and we extend our tool accordingly. For synchronous events, our tool automatically infers the data flow, and reports on inconsistencies due to circular data flow. We conduct two realistic case studies: a train control system (Sec.~\ref{index:example:train}) using indexed events, and a part of a nuclear shutdown system (Sec.~\ref{sync:example:nop}) using synchronous events. 

\smallskip

\noindent\textit{Resources}. Complete details of the two case studies are included in an extended report~\cite{OWH14}, which also contains more case studies of cyber physical systems (i.e., a mutual exclusion protocol, and function blocks from the IEC~61131 Standard for programmable logic controllers) that can be specified using the new notations. Complete TTM listings of the case studies are available at: \url{https://wiki.eecs.yorku.ca/project/ttm/index_sync_evt}.

\section{Semantics for Indexed and Synchronous Events}\label{semantics}
\newcommand{\state}{\text{STATE}}
\newcommand{\timers}{\text{TIMER}}
\newcommand{\T}{{\mathbf{T}}}

We extend the one-step operational semantics of TTMs reported in~\cite{Ostroff2014} to support both indexed events (Sec.~\ref{sec:semantics:op_sem}) and synchronous events (Sec.~\ref{sec:semantics:composition}). The extensions involve redefining: 1) the abstract syntax of events which affects the rules of transitions and scheduling; and 2) the rules of module compositions. We include the most relevant details to present these extensions, while the complete account of the new semantics is included in an extended report~\cite[Sec. 6]{OWH14}. 

\smallskip

\npar{Abstract Syntax: Introducing Fair and Demonic Event Indices}\label{sec:semantics:abstract_syntax}

We define the abstract syntax of a TTM module instance ${\cal M}$ as a 5-tuple $(V, s_0, T, t_0, E)$ where 1) $V$ is a set of local or interface variables; 2) $T$ is a set of timers; 3) $E$ is a set of state-changing events; 4) $s_0 \in \text{STATE}$ is the initial state ($\text{STATE} ~\, \triangleq ~ V \rightarrow \text{VALUE}$); and 5) $t_0 \in \timers$ is the initial timer assignment ($\timers  ~\, \triangleq ~ T \rightarrow \nat$). We define $type \in T \tfun \pow(\nat)$ and $boundt \in T \tfun \nat$ for querying about, respectively, the type and upper bound of each timer. For example, if timer $t_1$ is declared as $t_1 : 0 .. 5$, then $boundt(t_1) = 5$ and $type(t_1) = \{0 .. 6\}$. Timers count up to one beyond the specified bound, and remain unchanged until they are started again. The figure below presents the generic form of a TTM event, where $V=\{v_1, v_2, v_3, \cdots\}$ and $T=\{t_1, t_2, t_3, t_4, \cdots\}$.

\smallskip

\noindent\begin{tabular}{l|l}
\begin{minipage}[t]{.48\linewidth}
\textbf{Concrete syntax of event} $e$:
\label{event:e}
\begin{ttmcodel}
  $event\_id$ ($x$ : fair $T_x$; $y$ : $T_y$) [$l$,$u$] just
   when $grd$
   start $t_1, t_2$
   stop $t_3, t_4$
   do $v_1 := exp_1$, 	
     if $condition$ then $v_2 := v_1' + exp_2$ 
     else skip fi, 
     $v_3 :: 1 .. 4$
   end 
\end{ttmcodel}
\end{minipage}
&
\begin{minipage}[t]{0.53\linewidth}
\textbf{Abstract syntax of the event} $e$:
\small
\begin{itemize}
\item $e.id \in \text{ID}$;
\item $e.f\_ind \subseteq \text{ID}$ \ ; \ $e.d\_ind \subseteq \text{ID} $ 
\item $e.d\_ind \triangleq e.f\_ind \bunion d\_ind $
\item $e.l \in \nat$; ~ $e.u \in \nat \bunion \{ \infty \} $
\item $\begin{array}{l}e.fair \\ \quad \in \{\text{spontaneous}, \text{just}, \text{compassionate}\}\end{array}$
\item $e.grd \in \text{STATE} \times \timers \tfun \Bool$;
\item $e.start \subseteq T$;
\item $e.stop \subseteq T$;
\item $e.action \in \state \times \timers \leftrightarrow \state$;  
\end{itemize}
\end{minipage}
\end{tabular}

\smallskip

We use a 10-tuple $(id, f\_ind, d\_ind, l, u, fair, grd, start, stop, action)$ to define the abstract syntax of an event $e$. We write $e.id$ for its identifier. Sets $e.f\_ind$ and $e.d\_ind$ contain, respectively, fair and demonic indices that can be referenced in the event. Its fairness assumption (i.e., $e.\mathit{fair}$), as discussed in Sec.~\ref{sec:semantics:op_sem}, filters out certain execution traces that will be considered in the model checking process. Its guard (i.e., $e.grd$) is a Boolean expression referencing state variables, timers, or its indices. An event $e$ must be taken between its lower time bound (LTB) $e.l$ and upper time bound (UTB) $e.u$, while its guard $e.grd$ remains true. The event action involves simultaneous assignments to $v_1, v_2, \cdots$. We write $v_3 :: 1 .. 4$ for a demonic (non-deterministic) assignment to $v_3$ from a finite range. Therefore, its state effect is a relation $e.action$ on state variables and timers. On the RHS of an assignment $y:=x$, the state variable $x$ may be ``primed'' ($x'$) or ``unprimed''. A primed variable refers to its value at the \emph{next} state, or its current-state value if it is unprimed. The use of primed variables in expressions allows for more expressive descriptions of state changes, especially when combined with the use of synchronous events (Sec.~\ref{sec:semantics:composition}).

\smallskip

\npar{Operational Semantics}\label{sec:semantics:op_sem}

Given a TTM module instance ${\cal M}$, an LTS (Labelled Transition System) is a 4-tuple ${\cal L} = (\Pi, \pi_0, \T, \tfun)$ where 1) $\Pi$ is a set of system configurations; 2) $\pi_0 \in \Pi$ is an initial configuration; 3) $\T$ is a set of transitions names (defined below); and 4) $\tfun \subseteq \Pi \cprod \T \cprod \Pi$ is a transition relation.

We define $E_{id}$ as the set of event transition names, and $E_{fair}$ as the set of transition name prefixes, excluding values of demonic indices (i.e., including values of fair indices): $E_{id} ~\triangleq~ \{ e,m ~|~ e \in E \,\land\, m \in e.f\_ind \tfun \text{VALUE}   ~\spot~ (e.id, m) \}$. On the one hand, we use $e(x)$ to denote the (external) transition name of event $e$ with $x$, the values of its fair indices. On the other hand, when referring to the occurrence of $e$, in an LTL formula for instance, we use $e(x,y)$ to include $y$, the values of its demonic indices; otherwise, values of demonic indices are treated as internal non-deterministic choice within the event.

A configuration $\pi \in \Pi$ is defined by a 6-tuple $(s,t,m,c,x,p)$, where:

\smallskip
\noindent$\bullet$ $s \in \state$ is a value assignment for all the variables of the system. The state can be read and changed by any transition corresponding to an event in $E$.

\smallskip
\noindent$\bullet$ $t \in  \timers$ is a timer valuation function. Event transitions may start, stop, and read timers. A $tick$ transition representing a global clock changes the timers. 

\smallskip
\noindent$\bullet$  $m \in T \tfun \Bool$ records the status of monotonicity of each timer. Suppose event $e_1$ starts $t_1$, then we may specify that a predicate $p$ becomes true within 4 ticks after $e_1$'s occurrence. However, other events might stop or restart $t_1$ before $p$ is satisfied, making $t_1$ not in sync with the global clock. The expression $m(t_1)$ (monotonicity of timer $t_1$) holds in any state where $t_1$ is not stopped or reset. 

\smallskip
\noindent$\bullet$  $c \in E_{id} \tfun \nat \bunion \{ -1 \}$ is a value assignment for a clock implicitly associated with each event.  These clocks are used to decide whether an event has been enabled for long enough ($c(e.id, x) \geq e.l$) and whether it is urgent ($c(e.id, x) = e.u$). 

\smallskip
\noindent$\bullet$  $x \in E_{id} \bunion \{ \bot \}$ provides a sequencing mechanism: each transition $e$ is immediately preceded by a transition $e\#$ to update the monotonicity record $m$. 

\smallskip
\noindent$\bullet$  $p \in E_{id} \bunion \{ tick, \bot \}$ holds the name of the last event to be taken at each configuration. It is $\bot$ in the initial configuration. It allows us to refer to events in LTL formula, to state that they have just occurred. 

\smallskip

We focus on components $s$ and $c$ that are affected the most by fair and demonic indices, whereas components $t$, $m$, and $x$, as to how the monotonicity status of timers is maintained, are less relevant and included in~\cite[Sec. 6]{OWH14}.

Given a flattened module instance $\mathcal{M}$, transitions of its corresponding LTS are given as ${\T} = E_{id} \bunion E\# \bunion \{ tick \}$, where $E\# ~\triangleq~ \{ e \in E_{id} \spot e\# \}$ is the set of monotonicity-breaking transitions as mentioned above. Explicit timers and event (lower and upper) time bounds are described with respect to this tick transition. We define the enabling condition of event $e \in E$ with fair index $x$ and demonic index $y$ as when its guard is satisfied, and when its implicit clock is in-between its specified bounds: $(~e.en(x) ~\triangleq~ (\exists y \spot e.grd(x,y)) ~\land~ e.l \le c(e.id, x) \le e.u~)$.

The initial configuration is defined as $\pi_0 = (s_0,t_0, m_0, c_0, \bot,\bot)$, where $s_0$ and $t_0$ come from the abstract (Sec.~\ref{sec:semantics:abstract_syntax}). The value of each event $e_i$'s implicit clock depends on its guard being satisfied initially. More precisely, $c_0(e_i.id, x)$ equals 0 (the clock starts) if $(s_0,t_0) \models (\exists y \spot e_i.grd(x,y))$\footnote{If a state-formula $q$ holds in a configuration $\pi$, then we write $\pi \vDash q$. For formulas such as guards which do not depend on all components of a configuration, we drop some of its components on the left of $\models$, as in $(s_0, t_0) \models e.grd(x,y)$.
}; otherwise, it equals -1.

An execution $\sigma$ of the LTS {\cal L} is an infinite sequence $\pi_0 \stackrel{\tau_1}{\tfun} \pi_1 \stackrel{\tau_2}{\tfun} \pi_2 \tfun \cdots$, alternating between configurations $\pi_i \in \Pi$ and transitions $\tau_i \in \T$. Below, we provide constraints on each one-step relation ($\pi \stackrel{e}{\tfun} \pi'$) in an execution. If an execution $\sigma$ satisfies all these constraints then we call  $\sigma$ a \emph{legal} execution. To characterize the complete behaviour of ${\cal L}$, we let $\Sigma_{\cal L}$ denote the set of all its legal executions. Given a temporal logic property $\varphi$ and an LTS ${\cal L}$, we write ${\cal L} \vDash \varphi$ iff $\forall \sigma \in \Sigma_{\cal L} \spot \sigma \vDash \varphi$. There are two possible transition steps (event $e(x)$ and $tick$):
\begin{eqnarray}
(s,t,m,c,e(x), p) \stackrel{e(x)}{\tfun} (s',t',m',c',\bot, e(x)) \label{e}\\
(s,t,m,c,\bot, p) \stackrel{tick}{\tfun} (s,t',m',c',\bot, tick)\label{tick}
\end{eqnarray}

\noindent\textbf{Taking \emph{e}} The transition $e(x)$ specified in Eq.~\ref{e} is taken only if  the $x$-component of the configuration is $e$ (meaning that $e$\# was just taken, so $e$ is the only event allowed to be taken) and $(s,t,c) \vDash e.en(x)$. The component $s'$ of the \emph{next} configuration in an execution is determined non-deterministically by $e.action(x,y)$, which is a relation as demonic indices or assignments may be used. Consequently, any next configuration that satisfies the relation can be part of a valid execution, i.e., $s'$ is only constrained by $(s,t,s') \in e.action(x,y)$. The following function tables specify the updates to $c$ upon occurrence of transition $e(x)$.

\smallskip 

\begin{tabular}{|c|c||c|}
\hline
	\multicolumn{2}{|c||}{For each event $e_i\in E$, $x \in e_i.f\_ind \tfun \text{VALUE}$} & $c'(e_i.id)$ \\
\hline 
\hline
	\multicolumn{2}{|l||}{$(s',t') \not\models (\exists y \spot e_i.grd(x,y))$} & -1 \\
\hline
	\multirow{2}{*}{$(s',t') \models (\exists y \spot e_i.grd(x,y))$} 
	& $(s,t) \models (\exists y \spot e_i.grd(x,y)) ~\land~ \neg e_i = e$
	& $c(e_i.id, x)$ \\
\cline{2-3}
	& $(s,t) \not\models (\exists y \spot e_i.grd(x,y)) ~\lor~ e_i = e$
	& 0 \\
\hline
\end{tabular}

\smallskip 

\noindent We start and stop the implicit clock of $e_i$ as a consequence of executing $e$, according to whether $e_i.grd$ just becomes or remains false (1st row), remains true (2nd row), or just becomes true (3rd row). Event $e_i$ is ready to be taken if it becomes enabled $e_i.l$ units after its guard becomes true. 

\smallskip

\noindent\textbf{Taking \emph{tick}} The tick transition specified in Eq.~\ref{tick} is taken only if the $x$-component of the configuration is $\bot$ (thus preventing $tick$ from intervening between any $e\#$ and $e$ pair) and if $\forall e \in E \spot c(e.id, x) < e.u$. 

\smallskip

\begin{tabular}{|c|c||c|}
\hline
	\multicolumn{2}{|c||}{For each event $e\in E$, $x \in e.f\_ind \tfun \text{VALUE}$} & $c'(e.id, x)$ \\
\hline 
\hline
	\multicolumn{2}{|l||}{$(s',t') \not\models (\exists y \spot e.grd(x,y))$} & -1 \\
\hline
	\multirow{2}{*}{$(s',t') \models (\exists y \spot e.grd(x,y))$} 
	& $(s,t) \not\models (\exists y \spot e.grd(x,y))$
	& 0 \\
\cline{2-3}
	& $(s,t) \models (\exists y \spot e.grd(x,y)) $
	& $c(e.id, x)+1$ \\
\hline
\end{tabular}

\smallskip

\smallskip

\noindent Thus, $tick$ increments timers and implicit clocks towards their upper bounds. 

\smallskip

\noindent\textbf{Scheduling} So far, we have constrained executions so that the state changes in controlled ways. However, to ensure that a given execution does not stop making progress, we need to assume fairness. The current TTM tool supports four possible scheduling assumptions.

\smallskip

\noindent\textit{1. Spontaneous event}. When no fairness keyword is given, and the UTB is given as \verb|*| or unspecified, then even when the event is enabled, it might never be taken.

\smallskip

\noindent\textit{2. Just event scheduling} (a.k.a. weak fairness~\cite{SunLDP09}). This is assumed when the event is declared with the keyword \verb|just| and when the upper time bound is * or unspecified. For any execution $\sigma \in \Sigma_\mathcal{L}$, if an event $e$ eventually becomes continuously enabled, then it occurs infinitely many times: $\sigma ~\vDash~ (\forall x \,\spot\, \lozenge \square e.en(x) \, \rightarrow \, \square \lozenge (\exists y \spot e(x,y)))$, where $x$ ranges over $e$'s fair indices and $y$ its demonic indices. 

This highlights the key distinction between fair and demonic indices. The fairness assumption guarantees that $e(x, \_)$ is treated fairly for every single value of $x$. For example, if $x$ is a process identifier, making it a fair index means that as long as it is active, each process is eventually given CPU time. In contrast, if $x$ is treated as a demonic index, then it is possible that infinitely often the same process will be given CPU time.

\smallskip

\noindent\textit{3. Compassionate event scheduling} (a.k.a. strong fairness~\cite{SunLDP09}). This is assumed when the event is declared with the keyword \verb|compassionate| and when the upper time bound is * or unspecified. For any execution $\sigma \in \Sigma_\mathcal{L}$, if an event $e$ becomes enabled infinitely many times, it has to occur infinitely many times. More precisely: $\sigma ~\vDash~ (\forall x \,\spot\,  \square \lozenge e.en(x) \, \rightarrow \, \square \lozenge (\exists y \spot e(x,y)) $.

\smallskip

\noindent\textit{4. Real-time event scheduling}. The finite UTB $e.u$ of the event $e$ is taken as a deadline: it has to occur within $u$ units of time after $e.grd$ becomes true or after the last occurrence of $e$. To achieve this effect, the event $e$ is treated as \verb|just|. Since $tick$ will not occur as long as $e$ is urgent (i.e.,~$e.c = e.u$), transition $e$ will be forced to occur (unless some other event occurs and disables it).

\smallskip

\npar{Semantics of Module Composition}\label{sec:semantics:composition} 

So far we have specified the semantics of individual module instances. However, the TTM notation includes a composition. The semantics of systems comprising many instances is defined through flattening, i.e. by providing a single instance which, by definition, has the same semantics as the whole system.

\smallskip

\noindent\textbf{Instantiation} When integrating modules in a system, they first have to be instantiated, meaning that the module interface variables must be linked to global variables of the system which it will be a part of. For example if we have a $Phil$ module (for philosopher) with two shared variables, $\mathit{left\_fork}$ and $\mathit{right\_fork}$, and two global fork variables $f1$ and $f2$, we may instantiate them as:

\begin{ttmcodel}
  instances p1 = Phil(share f1, share f2) ; p2 = Phil(share f2, share f1)  end
\end{ttmcodel}

\noindent Philosopher $p1$ is therefore equivalent to the module $Phil$ with its references to $\mathit{left\_fork}$ substituted by $f1$ and its references to $\mathit{right\_fork}$ substituted by $f2$.

\smallskip

\noindent\textbf{Composition} The composition $m1 || m2$ is an associative and commutative function on two module instances. To flatten the composition, we rename the local variables and events (by prepending the module instance name) so that they are system-wide unique. We then proceed to create the composite instance. Its local variables are the (disjoint) union of the local variables of the two instances. Its interface variables are the (possibly non-disjoint) union of the interface variables of both instances with their mode (\verb|in|, \verb|out|, \verb|share|) adjusted properly~\cite[Table 1, p. 38]{OWH14} (e.g., variable \verb|in x| in $m1$ and variable \verb|out x| in $m2$ result in an \verb|out| variable in the composite instance). 

The simplest case of composition results in the union of the set of events of both instances. However, events from separate instances can be executed synchronously. This can be specified using the notation of synchronous events. As an illustration, consider a case where the plant and controller act synchronously. 

\smallskip

\resizebox{.95\textwidth}{!}{\includegraphics[width=\textwidth]{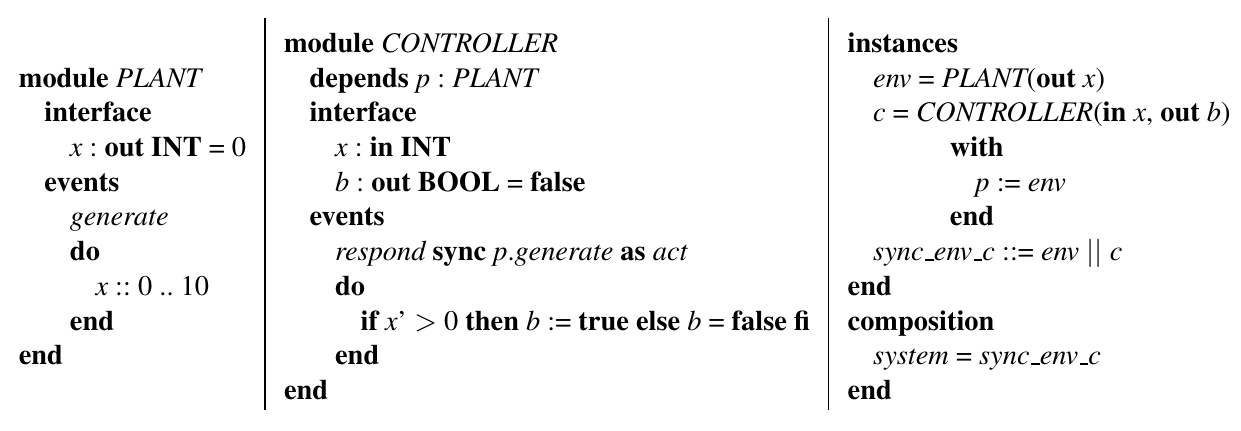}} 

\smallskip

\noindent We say module \var{CONTROLLER} depends on module \var{PLANT}. At the module level (e.g., \var{CTRL}), we use a \var{depends} clause to specify a list of instances that the current module depends on. At the event level (e.g., \var{respond}), we use a \var{sync} $\dots$ \var{as} $\dots$ clause to specify the list of events to be synchronized, qualified by names of the dependent instances (e.g., \var{p.generate}), and to rename the synchronized events with a new name (\var{act}). Actions of events that are involved in synchronization may reference the primed version of input variables to obtain their next-state values. For example, the \var{respond} event uses the next-state value of the input variable \var{x} (i.e., \var{x'}) to compute the next-state value of its output variable \var{b}. In creating an instance, we use a \var{with} $\dots$ \var{end} clause to bind all its dependent instances, if any. We use the \var{::=} operator to rename the synchronized instances (e.g., \var{sync\_env\_c}). As instances \var{env} and \var{c} are synchronized as the new instance \var{sync\_env\_c}, taking the event \var{sync\_env\_c.act} has the effect of updating, as one atomic step, the monitored variable \var{x} then controlled variable \var{b}. 

Specifying \var{depends} clauses (at the module level) and \var{sync} clauses (at the event level) results in one or more compound events whose actions are composed of those involved in the synchronization. We discuss the process of merging event actions below. For how event time bounds and fairness assumptions are merged in synchronization, refer to~\cite[p. 40]{OWH14}. 

The use of synchronous events results in three kinds of dependency graphs\footnote{Assume that \var{MOD} denotes the set of declared modules, \var{EVT} the set of declared events qualified by their containing modules, e.g., \var{PLANT.generate}, and \var{VAR} the set of interface and local variables}.

\smallskip

\noindent\textit{1.} The \textit{Module Dependency Graph} contains the set of vertices $\var{V} = \var{MOD}$, and the set of edges consisting of \var{(m$_1$,\ m$_2$)}, where module \var{m$_1$} depends on \var{m$_2$}. 

In each connected component of the module dependency graph, we construct a \emph{synchronous event set} (e.g., $\{ \var{PLANT.generate}, \var{CONTROLLER.respond } \}$) by including each event \var{e}, where \var{e} declares a \var{sync} clause, and all events under \var{e}'s \var{sync} clause. 

\smallskip

\noindent\textit{2.} An \textit{Event Dependency Graph} contains the set of vertices $\var{V} = \var{EVT}$, and the set of edges consisting of \var{(e$_1$,\ e$_2$)}, where \var{e$_1$} and \var{e$_2$} are in the same synchronous event set and \var{e$_2$} is declared under the \var{sync} clause of \var{e$_1$}.

\smallskip

\noindent\textit{3.} An \textit{Action Graph} is constructed from each synchronous event set. We write \var{VAR}$_s$ to denote variables that are involved in actions of events in a synchronous event set \var{s}. For each synchronous event set \var{s}, its corresponding action graph contains the set of vertices $\var{V} = \var{VAR}_s$, and the set of edges consisting of \var{(v$_1$,\ v$_2$)}, where the computation of \var{v$_1$}'s new (or next state) value depends on that of \var{v$_2$}. There are two cases to consider: 1) in an equation where \var{v$_2$} appears on the RHS and \var{v$_1$}' on the LHS (i.e., \var{v$_1$' = $\dots$ v$_2$ $\dots$}); and 2) in an assignment where \var{v$_2$} appears on the RHS and \var{v$_1$} on the LHS (i.e., \var{v$_1$ := $\dots$ v$_2$ $\dots$}).

We perform a topological sort on each action graph to calculate the order of variable assignments, from which we calculate a sequence of variable projections. The \emph{projection} for each variable \var{v} is a pair $(\var{v}, \var{act})$, where \var{act} is either an unconditional assignment (i.e., \var{v := exp}), or an conditional assignment (i.e., \var{\textbf{if} b$_1$ \textbf{then} v := exp$_1$ \textbf{elseif} b$_2$ \textbf{then} v := exp$_2$ $\dots$ \textbf{else} $\dots$}). The latter case is resulted from the fact that changes on \var{v} (either through assignments or the primed notation) occur inside nested if-statements. Finally, the produced sequence of variable projections is adopted as the action of the compound event.

To ensure consistency, the TTM tool reports an error when, e.g., one of the above graphs contains a cycle, or a flattened (or compound) event assigns multiples values to the same variable. 

\smallskip

\noindent\textbf{Iterated Composition}. Iterated composition allows us to compose an indexed set of similar instances. For example, in the case of a network of processes, we may specify the common process behaviour as a module once, and instantiate them from the set \var{PID} of process identifiers: $system = ||~pid : \textit{PID}~@~\textit{Process}(\textbf{in}~pid)$.




\section{Example: A Train Control System}\label{index:example:train}

We illustrate the use of TTM indexed events in a train control system. There are two reasons for using the indexed events. First, all trains entering and leaving the station share a common behaviour. Second, by declaring event indices (ranging over trains) as fair, we can assert that individual trains arriving at the station are guaranteed to depart, without being blocked indefinitely by other trains. 

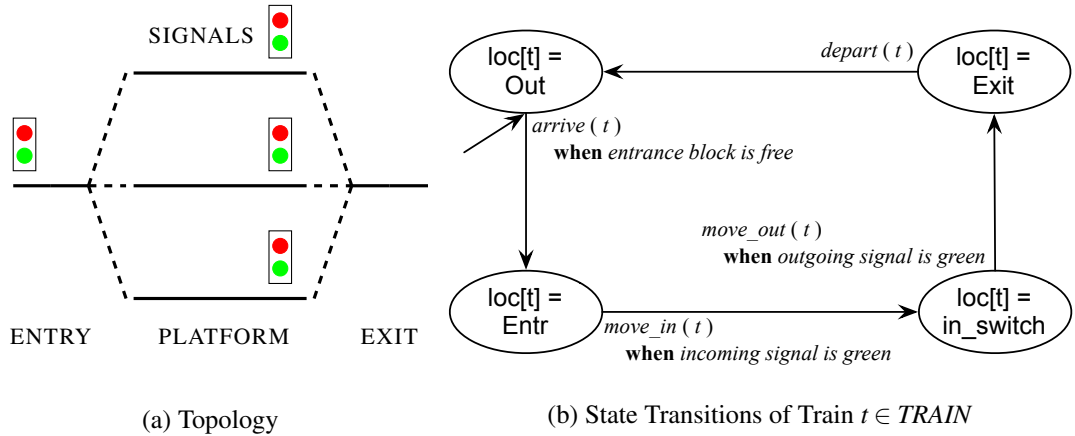
\begin{figure}[h]
\centering
\begin{subfigure}{.35\textwidth}
\begin{tikzpicture}[scale=1]
  \draw[very thick] (1,0) -- (1.5,0);
  \draw (1.5, -2) node{\textsc{entry}};

  \draw[very thick] (1.5,0) -- (2,0);
  \draw[very thick, dashed] (2,0) -- (2.5,1.5);
  \draw[very thick, dashed] (2,0) -- (2.5,-1.5);
  \draw[very thick, dashed] (2,0) -- (2.5,0);
  
  \draw[very thick] (2.6,0) -- (4.9,0);
  \draw[very thick] (2.6,1.5) -- (4.9,1.5);
  \draw[very thick] (2.6,-1.5) -- (4.9,-1.5);
  \draw (3.8, -2) node{\textsc{platform}};

  \draw[very thick] (5.5,0) -- (6,0);
  \draw[very thick, dashed] (5,1.5) -- (5.5,0);
  \draw[very thick, dashed] (5,0) -- (5.5,0);
  \draw[ very thick, dashed] (5,-1.5) -- (5.5,0);

  \draw[very thick] (6.0,0) -- (6.6,0);
  \draw (6, -2) node{\textsc{exit}};

  \draw (1, 0.2) rectangle +(0.3,0.7);
  \filldraw[red] (1.15, 0.7) circle (0.1);
  \filldraw[green] (1.15, 0.4) circle (0.1);

  \draw (4.4, 0.2) rectangle +(0.3,0.7);
  \filldraw[red] (4.55, 0.7) circle (0.1);
  \filldraw[green] (4.55, 0.4) circle (0.1);

  \draw (4.4, 1.7) rectangle +(0.3,0.7);
  \filldraw[red] (4.55, 2.2) circle (0.1);
  \filldraw[green] (4.55, 1.9) circle (0.1);
  \draw (3.5, 2) node{\textsc{signals}};

  \draw (4.4, -1.3) rectangle +(0.3,0.7);
  \filldraw[red] (4.55, -0.8) circle (0.1);
  \filldraw[green] (4.55, -1.1) circle (0.1);

\end{tikzpicture}
\caption{Topology}\label{fig:train:system:topology}
\end{subfigure}
\begin{subfigure}{.55\textwidth}
\includegraphics[width=\textwidth]{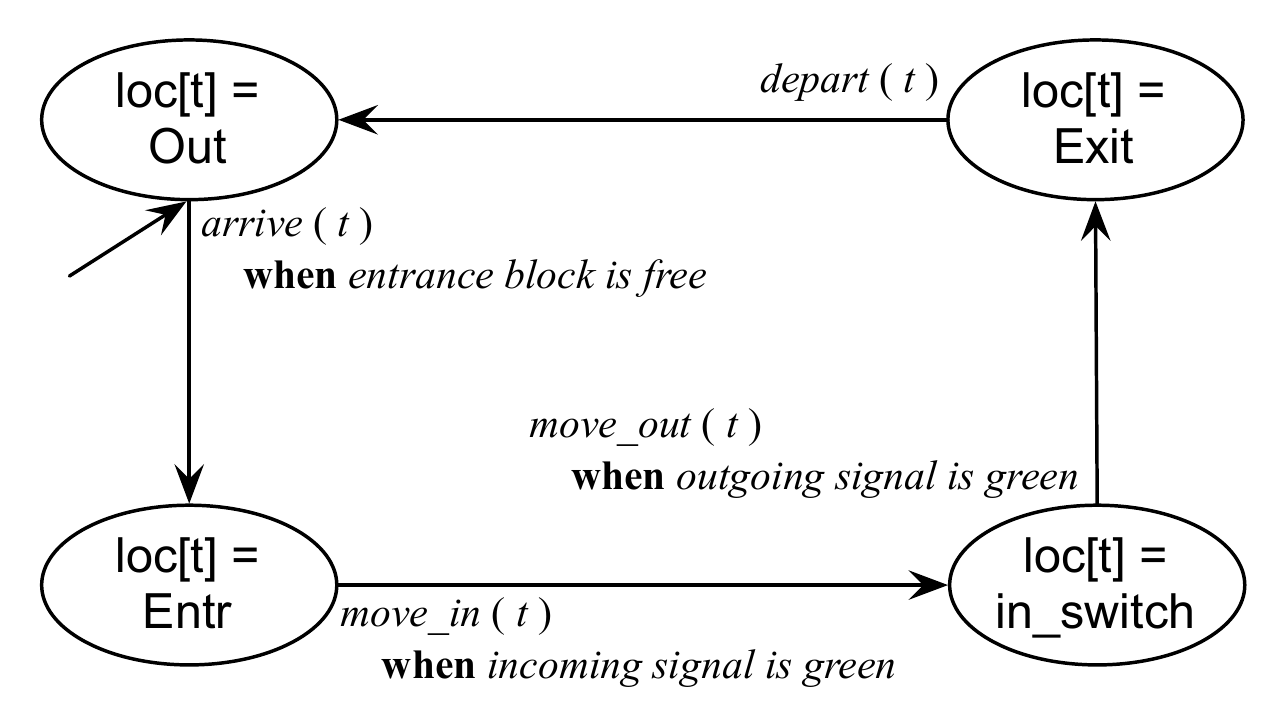}
\caption{State Transitions of Train $\var{t} \in \var{TRAIN}$}\label{fig:train:system:transition}
\end{subfigure}
\caption{A Train Control System}
\end{figure}

Fig.~\ref{fig:train:system:topology} shows the topology of the train control system~\cite{Hudon2013}. There is an entry block (\var{Entr}) and an exit block (\var{Exit}) on both ends of the station. Between the entry and exit blocks is a set \var{PLF} of special blocks called platforms. At most one train may stay at the entry or exit block at a time. On the entry bock, there is a signal \var{isgn} regulating the incoming train, depending on the availability of platforms. On each platform $\var{p} \in \var{PLF}$, there is a signal \var{osgn}[\var{p}] regulating the outgoing train, depending on the availability of the exit block. Fig.~\ref{fig:train:system:transition} illustrate the common behaviour of all trains. Each train is initially travelling outside the station. The train may first arrive at the entry block, provided that it is not occupied. When the signal \var{isgn} turns green, the train is directed via an in-switch to move in an available platform. For some train \var{t}, after it moved to platform \var{p}, it waits for the light signal of platform \var{p} to turn green and then moves away from \var{p} and onto the exit block. Then the train  may depart from the station. 

\begin{figure}[h]
\centering
\begin{subfigure}{.33\textwidth}\centering
\includegraphics[width=\textwidth]{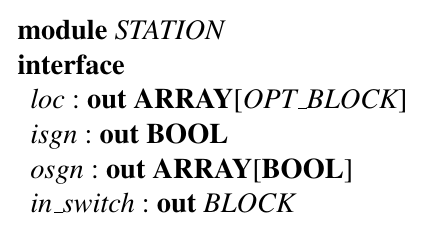}
\vspace{.2cm}
\caption{Abstract}\label{fig:train:interface:abstract}
\end{subfigure}
\begin{subfigure}{.57\textwidth}\centering
\includegraphics[width=\textwidth]{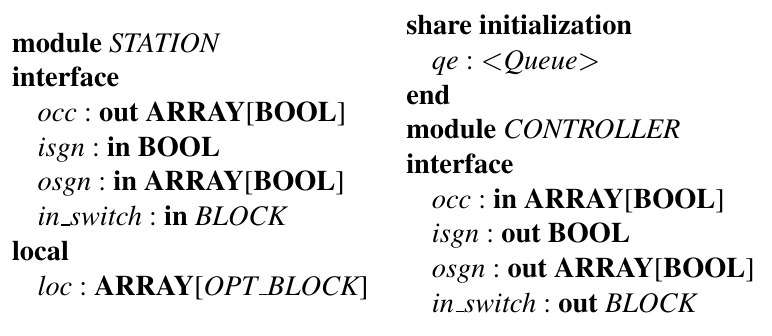}
\caption{Refined: Separate Station \& Controller Events}\label{fig:train:interface:refined}
\end{subfigure}
\caption{Train Control System in TTM: Interfaces}
\end{figure}

Trains must never collide in the train station. Also, once a train arrives, it should be eventually scheduled to depart from the station. 
{\small\begin{align}
& (\forall \var{t1}, \var{t2} : \var{TRAIN} \; \bullet \; (~\var{t1} \neq \var{t2} \land \var{loc}[\var{t1}] \neq \var{Out} \land \var{loc}[\var{t2}] \neq \var{Out}~) \Rightarrow~ \var{loc}[\var{t1}] \neq \var{loc}[\var{t2}]) \label{eq:train:safety} \\
& \Box(~ \var{loc}[t] = \var{Entr} \Rightarrow \Diamond (\var{loc}[t] = \var{Out}) ~) \label{eq:train:liveness1} 
\end{align}}
%
%
\noindent We consider two versions of TTM that satisfy both Eq.~\ref{eq:train:safety} and \ref{eq:train:liveness1}. Fig.~\ref{fig:train:interface:abstract} presents the TTM interface of an abstract version, where monitored and controlled variables are separated. As a result, the abstract version contains a single \var{STATION} module that: (a) owns all variables; and (b) mixes all events of train movement (e.g., event \var{move\_out} in Fig.~\ref{fig:train:move_in:abstract}) and of signal control (e.g., event \var{ctrl\_platform\_signal} in Fig.~\ref{fig:train:ctrl_event:abstract}). On the other hand, Fig.~\ref{fig:train:interface:refined} presents the interface of a refined version, which distinguishes between one monitored variable (i.e., \var{occ} for the set of occupied platforms) and three controlled variables (i.e, \var{isgn} for an incoming train, \var{in\_switch} for platform currently connected to the entrance block, and \var{osgn} for outgoing trains). Consistently, the behaviour of the controller and that of the trains are factored in separate events and placed in separate modules. The monitored variable (with modifier \var{\textbf{in}}) is owned by the \var{STATION} module and read-only for the \var{CONTROLLER} module. 

The refined version of TTM changes the representation of the data used by control events. In the abstract version (Fig.~\ref{fig:train:interface:abstract}), the array variable \var{loc} is used to map each train to its current location, constrained by type $\var{OPT\_BLOCK} \triangleq \{\var{Out}\} \cup \var{BLOCK}$ where $\var{BLOCK} \triangleq \{ \var{Entr},\ \var{Exit}\} \cup \var{PLF}$. All train events (e.g., \var{move\_out} in Fig.~\ref{fig:train:move_in:abstract}) are indexed with the set of trains and update their location accordingly (e.g., \var{loc}[\var{t}] \var{:= Exit}). All control events (e.g., \var{ctrl\_platform\_signal} in Fig.~\ref{fig:train:ctrl_event:abstract}) query the value of \var{loc} in their guards (e.g., we write \var{$!$(~$||$t: TRAIN @ loc\emph{[}t\emph{]} == Exit~)} to check that the exit block is not occupied). However, a more realistic station controller may monitor platforms in the station only, rather than all trains including those travelling elsewhere outside the station. Consequently, in the refined version (Fig.~\ref{fig:train:interface:refined}), by refactoring \var{loc} as a local variable in the \var{STATION} module (the environment), we hide it from the \var{CONTROLLER}. The controller then only has access to the monitored variable \var{occ} (i.e., the set of occupied platforms) which encodes a coarser grain of information than \var{loc} (i.e., locations of all trains). Using the new monitored variable \var{occ} simplifies guards of controller events (Fig.~\ref{fig:train:ctrl_event:refined}). 
Moreover, train events in the environment (e.g., Fig.~\ref{fig:train:move_in:refined}) updates both the local variable \var{loc} and the output variable \var{occ}. This raises the question of whether the \var{CONTROLLER} module accesses the monitored variable \var{occ} in a way consistent with the corresponding events in the abstract model. Therefore, we assert that a block is occupied if and only if it corresponds to the location of some train.

\begin{figure}[h]
\centering
\begin{subfigure}{.43\textwidth}
\begin{ttmcodel}
move_out(t : fair TRAIN) just
 when call(is_platform,loc[t]) && osgn[loc[t]]
 do loc[t] := Exit, osgn[loc[t]] := false end 
\end{ttmcodel}
\caption{Abstract Version}\label{fig:train:move_in:abstract}
\end{subfigure}
\begin{subfigure}{.55\textwidth}
\begin{ttmcodel}
move_out(t : fair TRAIN)[2, *] just
 when call(is_platform,loc[t]) && osgn[loc[t]]
 do loc[t] := Exit, occ[loc[t]] := false, occ[Exit] := true end 
\end{ttmcodel}
\caption{Refined Version}\label{fig:train:move_in:refined}
\end{subfigure}
\caption{Train Control System in TTM: the \var{move\_out} Event in Module \var{STATION}}
\end{figure}

The two versions of TTMs are different in scheduling the green signals that control the passage from the platforms to the exit block. While the abstract model is non-deterministic about the order in which trains gain access to the exit block, the concrete model specifies the order uniquely. The signals are controlled by event \var{ctrl\_platform\_signal}. In the abstract version (Fig.~\ref{fig:train:ctrl_event:abstract}), the event is indexed by the set of trains. When the exit block is not occupied, more than one train located at a platforms may be eligible to move on to the exit block. To satisfy Property~\ref{eq:train:liveness1}, we declare the index on trains as fair and adopt a strong fairness assumption (i.e., \var{compassionate}) on the controller event. That is, a train infinitely often qualified to leave the station does so eventually. However, such fairness assumption cannot be implemented efficiently. Consequently, in the refined version (Fig.~\ref{fig:train:ctrl_event:refined}), we use a C\# FIFO \var{Queue}\footnote{Using a C\# data object, implementation details of operations such as \var{Enqueue} are all encapsulated, resulting in a model simpler than one using a native TTM array.} to specify the order of train departure. The reduced non-determinism allows us to remove the fair index on trains and weaken the fairness assumption (i.e., the event becomes \var{just}).

\begin{figure}[h]
\begin{subfigure}{.56\textwidth}
\begin{ttmcodel}
ctrl_platform_signal(p : fair BLOCK) compassionate
 when call(is_platform, p)                              
   && (&&p : BLOCK @ call(is_platform, p) -> !osgn[p])  
   && !(||t: TRAIN @ loc[t] == Exit)                    
   && (||t: TRAIN @ loc[t] == p)                          
 do osgn[p] := true end
\end{ttmcodel}
\caption{Abstract Version in module \var{STATION}}\label{fig:train:ctrl_event:abstract}
\end{subfigure}
\begin{subfigure}{.44\textwidth}
\begin{ttmcodel}
ctrl_platform_signal just
 when qe.Count() != 0   
   && !osgn[qe.First()] 
   && !occ[Exit] 
   && occ[qe.First()]     
 do osgn[qe.First()] := true end 
\end{ttmcodel}
\caption{Refined Version in module \var{CONTROLLER}}\label{fig:train:ctrl_event:refined}
\end{subfigure}
\caption{Train Control System in TTM: Controller Events}
\end{figure}






\section{Example: Tabular Requirement of a Nuclear Shutdown System}\label{sync:example:nop}

We illustrate the use of synchronous events on parts of the software requirements of a shutdown system for the Darlington Nuclear Generating Station. We present two versions of the system. The first version presents a high-level requirements~\cite{Wassyng2006} where the controller responds instantaneously to environment changes. We synchronize the environment and controller events to model such instantaneity, and check it via an invariant property. The refined version illustrates how the response allowance~\cite{Wassyng2005} can be incorporated as event time bounds (i.e, the controller responds fast enough to environment changes). We decouple the controller from the environment, and check its response via a real-time liveness property.


Requirements of the shutdown system are described mathematically using tabular expressions (a.k.a. function tables)~\cite{Janicki1997}. Figure~\ref{fig:nop:table} exemplifies tabular requirements for two units: Neutron OverPower (NOP) Parameter Trip (Figure~\ref{fig:nop:table:controller}) and Sensor Trips (Figure~\ref{fig:nop:table:sensor}). In the first column, rows are Boolean conditions on monitored variables (i.e., input stimuli). In the second column, the first row names a controlled variable (i.e., output response); the remaining rows specify a value for that controlled variable. We use the formalism of tabular expressions to check the completeness (i.e., no missing cases from input conditions) and the disjointness (i.e., no input conditions satisfied simultaneously) of our requirements~\cite{Janicki1997}.  

\begin{figure}[h]
\centering
\begin{subfigure}{\textwidth}
\centering
{\small \begin{tabular}{||c|c||}
\multicolumn{1}{c}{}	                     & \multicolumn{1}{c}{\var{Result}} \\
\cline{2-2} 
\multicolumn{1}{c||}{\var{Condition}} & \var{c\_NOPparmtrip} \\
\hline \hline
$\exists \var{i} \in 0 \upto 17 \; \bullet \; \var{f\_NOPsentrip}[\var{i}] = \var{e\_Trip}$ & \var{e\_Trip} \\
\hline
$\forall \var{i} \in 0 \upto 17 \; \bullet \; \var{f\_NOPsentrip}[\var{i}] = \var{e\_NotTrip}$ & \var{e\_NotTrip} \\
\hline
\end{tabular}}
\caption{Function Table for NOP Controller}\label{fig:nop:table:controller}
\end{subfigure}

\begin{subfigure}{\textwidth}
\centering
{\small \begin{tabular}{||c|c||}
\multicolumn{1}{c}{}	                     & \multicolumn{1}{c}{\var{Result}} \\
\cline{2-2} 
\multicolumn{1}{c||}{\var{Condition}} & \var{f\_NOPsentrip}[\var{i}] \\
\hline \hline
\var{calibrated\_nop\_signal}[\var{i}] $\geq$ \var{f\_NOPsp} & \var{e\_Trip} \\
\hline
\var{f\_NOPsp} $-$ \var{k\_NOPhys} $<$ \var{calibrated\_nop\_signal}[\var{i}] $<$ \var{f\_NOPsp} & (\var{f\_NOPsentrip}[\var{i}])$_{-1}$ \\
\hline
\var{calibrated\_nop\_signal}[\var{i}] $\leq$ \var{f\_NOPsp} $-$ \var{k\_NOPhys} & \var{e\_NotTrip} \\
\hline
\end{tabular}}
\caption{Function Table for NOP sensor \var{i}, $\var{i} \in 0 \upto 17$ (monitoring \var{calibrated\_nop\_signal}[\var{i}])}\label{fig:nop:table:sensor}
\end{subfigure}


\caption{Tabular Requirement for the Neutron Overpower (NOP) Trip Unit}\label{fig:nop:table}
\end{figure}

The NOP Parameter Trip unit (the NOP controller) depends on 18 instances of the Sensor Trip units (the NOP sensors). There are two monitored variables for each NOP sensor \var{i}: (1) a floating-point calibrated NOP signal value \var{calibrated\_nop\_signal}[\var{i}]; and (2) a floating-point set point value \var{f\_NOPsp}. The monitored signal is bounded by the two pre-set constants \var{k\_NOPLoLimit} and \var{k\_NOPHILimit}. The monitored set point can be one of the four constants: \var{k\_NOPLPsp} (low-power mode), \var{k\_NOPAbn2sp} (abnormal mode~2), \var{k\_NOPAbn1sp} (abnormal mode~1), and \var{k\_NOPnormsp} (normal mode). 

Each sensor \var{i} determines if the monitored signal goes above a safety range (i.e., $\geq \var{f\_NOPsp}$), in which case it trips by setting the function variable \var{f\_NOPsentrip}[\var{i}] to \var{e\_Trip}. To prevent the value of \var{f\_NOPsentrip} from alternating too often due to signal oscillation, a hysteresis region (or dead band) with constant size \var{k\_NOPhys} is created. The hysteresis region $(\var{f\_NOPsp} - \var{k\_NOPhys},\; \var{f\_NOPsp})$ is an open interval. When the monitored signal falls within this region, then the new value of \var{f\_NOPsentrip} remains as that in the previous state, denoted as $\var{f\_NOPsentrip}_{-1}$. The NOP controller is responsible for setting the controlled variable \var{c\_NOPparmtrip}, based on values of \var{f\_NOPsentrip}[\var{i}] from all its dependant sensors. If there is at least one sensor that trips, then the NOP parameter trips by setting $c\_NOPparmtrip$ to \var{e\_Trip}.

According to the requirements, the system is initialized in a conservative manner. Each calibrated NOP signal is set to its low limit \var{k\_NOPLoLimit}, but each \var{f\_NOPsentrip}[\var{i}] for sensor \var{i} and the controlled variable \var{c\_NOPparmtrip} are all set to \var{e\_Trip}. As we will see in our specification below (i.e., Equation~\ref{eq:nop:spec:system}), to ensure that the system satisfies the tabular specification in Figure~\ref{fig:nop:table}, the NOP controller must have completed its very first response (denote as predicate $\lnot\var{init\_response}$). 

The requirements model in Figure~\ref{fig:nop:table} uses a finite state machine, with an arbitrarily small clock tick, that describes an idealized behaviour. At each time tick \var{t}, monitored and controlled variables are updated instantaneously. State data such as $\var{f\_NOPsentrip}_{-1}$ are stored and used for the next state. However, to make such requirements implementable, some allowance on the controller's response must be provided~\cite{Wassyng2005}. As a result, we present two versions of the NOP system in TTM: (1) an abstract version with plant and controller taking synchronized actions; and (2) a refined version with the response allowance incorporated as time bounds of the environment and controller events. The refined version allows us to assert timed response properties (e.g., once the monitored signal goes above the safety range, the controller trips within 2 ticks of the clock).

\smallskip 

\textit{Abstraction of Input Signal Values.} The TTM tool, like other model checking tools, cannot handle the real-valued monitored variables \var{f\_NOPsp} and \var{calibrated\_nop\_signal}[\var{i}]. Instead, based on the given constants mentioned above, we partition the infinite domains of these two monitored variables into disjoint intervals. First, the four possible constant values for \var{f\_NOPsp} have a fixed order and are bounded by constant low and high limits of the calibrated NOP signal. More precisely, we have 6 boundary cases to consider: {\small $\var{k\_NOPLoLimit} < \var{k\_NOPLPsp} < \var{k\_NOPAbn2sp} < \var{k\_NOPAbn1sp} < \var{k\_NOPnormsp} < \var{k\_NOPHiLimit}$}. Second, each of the four possible set points has an associated hysteresis band, whose lower boundary is calculated by subtracting the constant band size \var{k\_NOPhys}, resulting in 4 additional boundaries\footnote{Value of (a) is still greater than \textit{k\_NOPLoLimit}, and similarly value of (d) is still smaller than \textit{k\_NOPHiLimit}.} to consider: (a) $\var{k\_NOPLPsp} - \var{k\_NOPhys}$; (b) $\var{k\_NOPAbn2sp} - \var{k\_NOPhys}$; (c) $\var{k\_NOPAbn1sp} - \var{k\_NOPhys}$; and (d) $\var{k\_NOPnormsp} - \var{k\_NOPhys}$. Consequently, we have 10 boundary cases and 9 in-between cases (e.g., $\var{k\_NOPLoLimit} < \var{signal} < \var{k\_NOPLPsp}$) to consider. Accordingly, we construct a finite integer set \var{cal\_nop} that covers all the 19 intervals. 

For the purpose of modelling and verifying the NOP controller and sensors in TTM, we parameterize the system by a positive integer \var{N} denoting the number of dependant sensors.

\smallskip

\begin{figure}[h]
	\centering
\includegraphics[width=.85\textwidth]{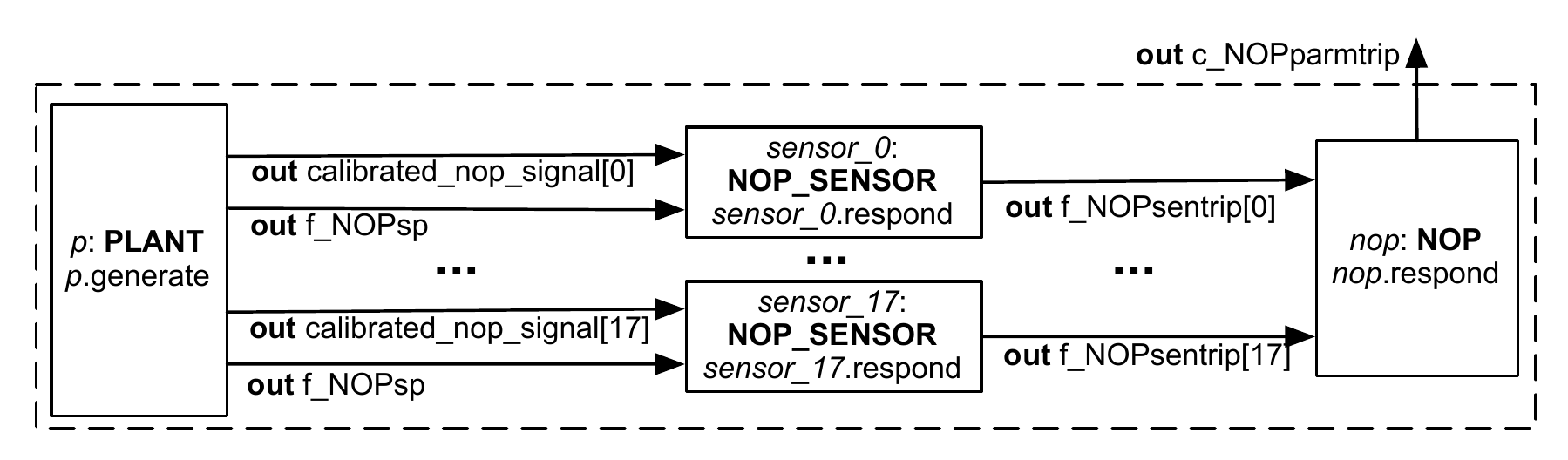}
\caption{Neutron Overpower (NOP): Abstract Version -- Synchronized Plant and Controller}\label{fig:nop:struc:abstract}
\end{figure}

\noindent\textbf{Version 1: Synchronizing Plant and Controller.} We first present an abstract version of the model that couples the NOP controller and its plant by executing their actions synchronously. Figure~\ref{fig:nop:struc:abstract} illustrates the structure of synchronization. The dashed box in Figure~\ref{fig:nop:struc:abstract} indicates the set of synchronized modules instances: plant \var{p}, controller \var{nop}, and 18 sensors \var{sensor\_i} ($i \in 0 \upto 17$). 

\begin{figure}[h]
\includegraphics[width=\textwidth]{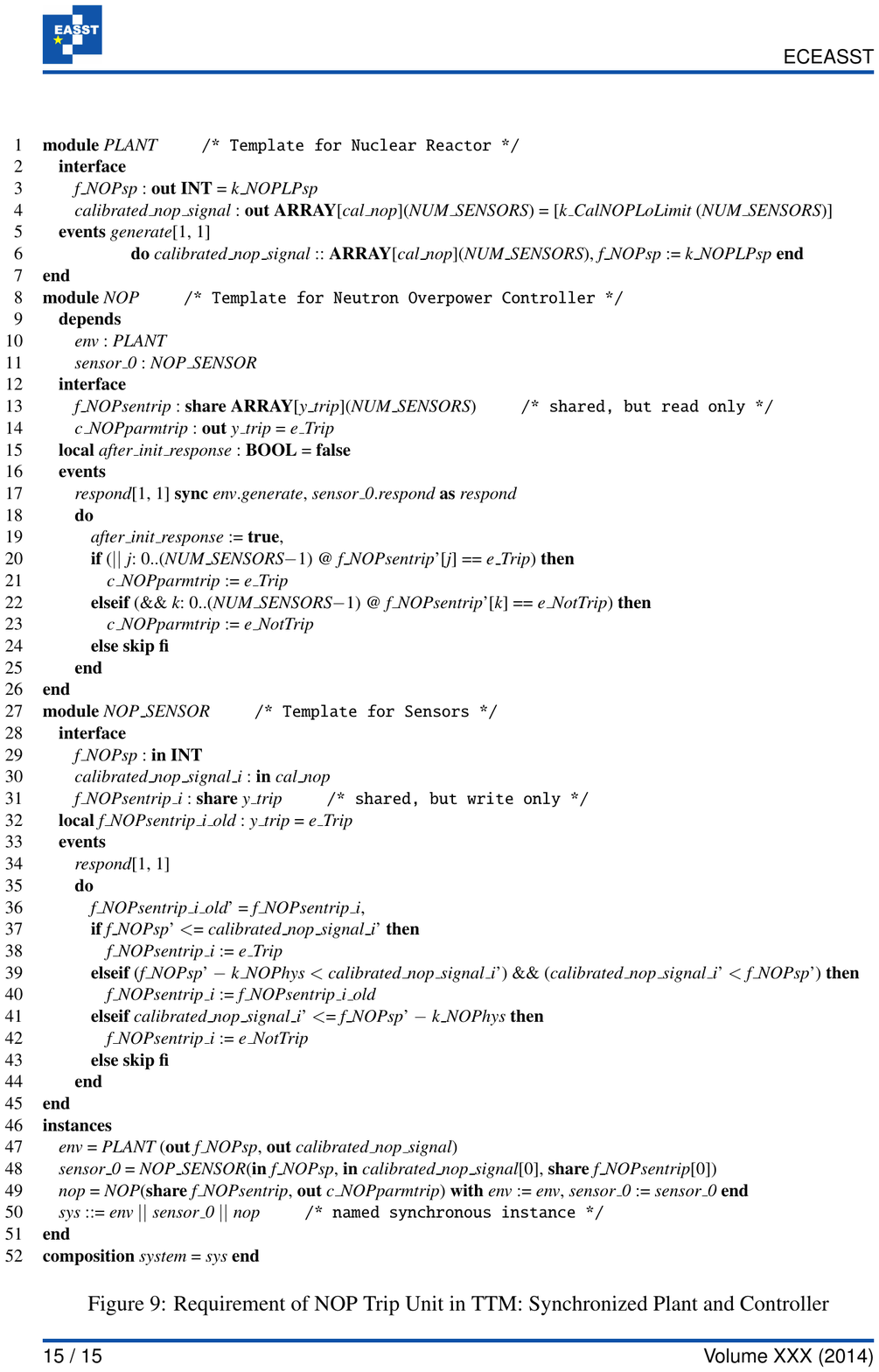}
\caption{Requirement of NOP in TTM: Synchronized Plant and Controller}\label{fig:nop:ttm}
\end{figure}

Figure~\ref{fig:nop:ttm} (p.~\pageref{fig:nop:ttm}) presents the complete\footnote{For clarity, we present a version with one monitoring sensor. The full version with 18 sensors involves just declaring and instantiating additional dependent sensors. We also exclude definitions of constants and assertions.} TTM listing of the NOP unit as described above. The \var{generate} event of the plant non-deterministically updates the value of a global array that is shared with sensors attached to the NOP controller. The update is performed via the demonic assignment \var{calibrated\_nop\_signals :: \textbf{ARRAY}}[\var{cal\_nop}](\var{N}) (Lines 5 -- 6). The NOP controller module (Lines 8 -- 26) depends on two module instances (Lines 9--11). First, the controller depends on a plant \var{p} that generates an array of calibrated NOP signals (specified by the \textbf{out} array argument \var{calibrated\_nop\_signal} at Lines 4 and 47). Second, the controller depends on a sensor \var{sensor\_0} that monitors a particular signal value (specified by the \textbf{in} argument \var{calibrated\_nop\_signal}[0] at Lines 30 and 48) and provides feedback (specified by the \textbf{share} argument \var{f\_NOPsentrip}[0] at Line 31 and 48) for the central NOP controller to make a final decision (specified by the \textbf{out} argument \var{c\_NOPparmtrip} at Lines 14 and 49). 

Actions of the \var{respond} events of the NOP controller (Lines 19 -- 24) and of its dependent sensors (Lines 36 -- 43) correspond to the tabular requirements (Figure~\ref{fig:nop:table:controller} and Figure~\ref{fig:nop:table:sensor}, respectively). We use primed variables in these actions to specify the intended flow of actions. Actions of the NOP sensor reference \var{f\_NOP'} and \var{calibrated\_nop\_signal\_i'} (Lines 37, 39, and 41) to indicate, that only after the instance \var{p} (in the same synchronous set) has written to these two variables can they be used to calculate the new value of \var{f\_NOPsentrip}[\var{i}]. Similarly, actions of the NOP controller reference \var{f\_NOPsentrip'}[\var{j}] (Lines 20 and 22) to indicate, that only after all sensor instances have written to this array can it be used to calculate the new value of \var{c\_NOPparmtrip}. 

We require that the \var{respond} event of the NOP controller, the \var{respond} events of its dependent sensors, and the \var{generate} event of the plant, are always executed synchronously (as a single transition). In declaring the controller event \var{respond}, we use a \var{\textbf{sync} $\dots$ \textbf{as} $\dots$} clause to specify the events to be included in the synchronous set. When instantiating the NOP controller, we use a \var{\textbf{with} $\dots$ \textbf{end}} clause to bind its dependent plant and sensor instances (Line 49). Finally, we rename the synchronized plant, controller, and sensor instances for references in assertions (Line 50).

We check two invariant properties on this abstract version of NOP. First, as all dependent sensors have written to the shared array \var{f\_NOPsentrip}, the NOP controller responds instantaneously.

{\small \begin{align}
\Box \left(
\begin{array}{cl}
      & (~\exists \var{i} : 0 \upto \var{N} \; \bullet \; \var{f\_NOPsentrip}[\var{i}] = \var{e\_Trip}~) 
        \Rightarrow \var{c\_NOPparmtrip} = \var{e\_Trip} \\
\land & (~\forall \var{i} : 0 \upto \var{N} \; \bullet \; \var{f\_NOPsentrip}[\var{i}] = \var{e\_NotTrip}~) 
        \Rightarrow \var{c\_NOPparmtrip} = \var{e\_NotTrip}
\end{array} \right)
\label{eq:nop:spec:controller}
\end{align}}

\noindent Second, since all actions of the plant, the NOP controller, and sensors are synchronized together, we can assert that the controlled variable \var{c\_NOPparmtrip} is updated as soon as the plant has updated the two monitored variables \var{f\_NOPsp} and \var{f\_NOPsentrip}.

{\small \begin{align}
\Box~\left( 
\begin{array}{l}
\left( 
\begin{array}{cl}
	  & \lnot~\var{init\_response} \\
\land & \var{f\_NOPsp} = \var{k\_NOPLPsp} \\
\land & \var{k\_NOPLPsp}\leq \var{calibrated\_nop\_signal}[0] \leq \var{k\_CalNOPHiLimit}
\end{array}
\right) \\ 
\qquad \qquad \Rightarrow \var{c\_NOPparmtrip} = \var{e\_Trip} 
\end{array} 
\right)
\label{eq:nop:spec:system}
\end{align}}

However, the satisfaction of Equation~\ref{eq:nop:spec:system} is an idealized behaviour without the realistic concern of some allowance on the controller's response~\cite{Wassyng2005}. That is, we shall instead allow the state predicate $\var{c\_NOPparmtrip} = \var{e\_Trip}$ to be established within a bounded delay. 

\smallskip

\begin{figure}[h]
	\centering
\includegraphics[width=.82\textwidth]{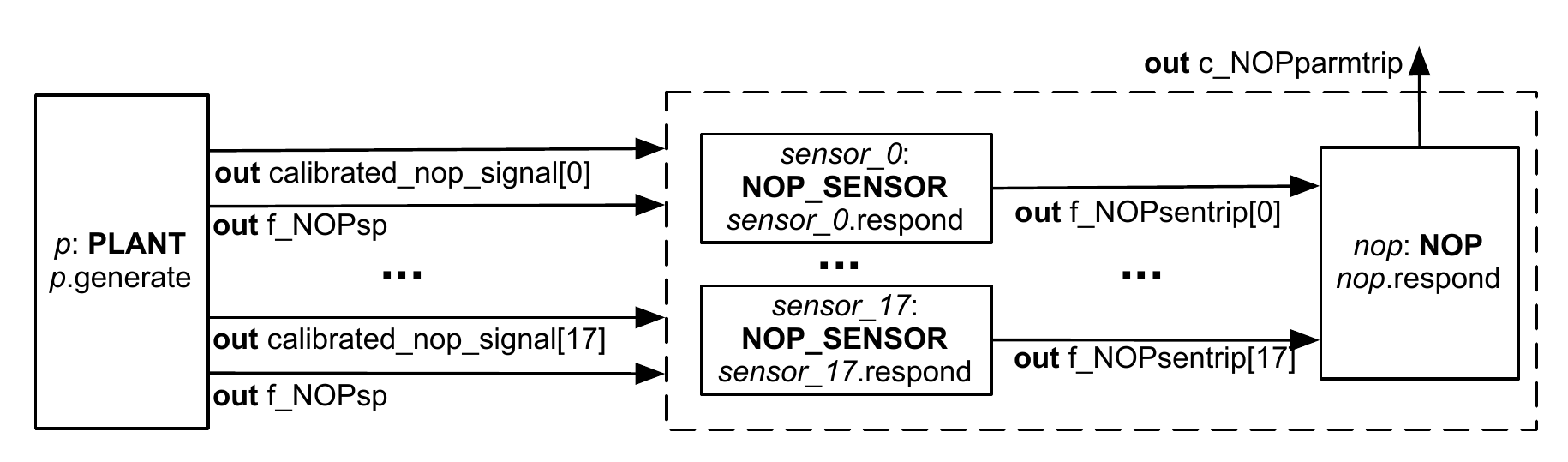}
\caption{Neutron Overpower (NOP): Refined Version -- Separate Plant and Controller}\label{fig:nop:struc:refined}
\end{figure}

\noindent\textbf{Version 2: Separating Plant and Controller.} We refine the TTM of NOP in Figure~\ref{fig:nop:ttm} by decoupling actions of the controller\footnote{In the NOP controller, actions of the NOP parameter trip unit and sensor units remain synchronized.} and its plant. Figure~\ref{fig:nop:struc:refined} illustrates the refined structure of synchronization: the plant instance \var{p} is no longer synchronized with the controller. Consequently, the plant event \var{generate} and the synchronous controller event \var{respond} are interleaved. 

The resulting system would fail to satisfy Equation~\ref{eq:nop:spec:system}, as we introduce some allowance on the response time (termed \emph{response allowance} in \cite{Wassyng2005}) of the NOP controller to environment changes. 
On the other hand, as we still consider the controller's response actions, once initiated, take effect instantaneously, the resulting system should still satisfy Equation~\ref{eq:nop:spec:controller}. 

We apply the following changes to produce the refined TTM (Figure~\ref{fig:nop:ttm}). First, in module \var{PLANT}, we revise time bounds of the \var{generate} event to $[2,\ *]$, which encodes the assumption that the controller (whose \var{respond} event has time bounds $[1,\ 1]$) responds fast enough to the environment changes. Second, in module \var{NOP}, we remove the declaration of \var{p : PLANT} as a dependent instance (Line 10). We also remove the declaration of \var{p.generate} as an event to be synchronized with the \var{respond} event (Line 17). Third, in creating the instance \var{nop} of module \var{NOP}, as it no longer depends on a \var{PLANT} instance, we remove the binding statement (Line 49), i.e., \var{env := env}. Fourth, in renaming the synchronous instance, we remove the plant instance (Line 50), i.e., \var{controller ::= sensor\_0 $\mathop{||}$ nop}. Finally, we add the plant instance into the composition (Line 52), i.e., \var{system = env $\mathop{||}$ controller}. 

By declaring a timer \var{t} and adding a \var{\textbf{start} t} clause to the \var{generate} event in module \var{PLANT} (Line 6), we can satisfy the following real-time response property:

{\small \begin{align}
\Box~\left( 
\begin{array}{l}
\left( 
\begin{array}{cl}
	  & \var{f\_NOPsp} = \var{k\_NOPLPsp} \\
\land & \var{k\_NOPLPsp}\leq \var{calibrated\_nop\_signal}[0] \leq \var{k\_CalNOPHiLimit} \\
\land & t = 0
\end{array}
\right) \\ 
\qquad \qquad \Rightarrow \var{mono}(\var{t}) \; \; \mathbf{U} \; \; (~\var{c\_NOPparmtrip} = \var{e\_Trip} \land \var{t} < 2~)
\end{array} 
\right)
\label{eq:nop:spec:response:liveness}
\end{align}}

\noindent As soon as the set point value and monitored signal value are updated by the plant, the controller produces the proper response within two ticks of the clock. Before the controller responds, timer \var{t} must not be interrupted (i.e., reset by other events), so as not to provide an inaccurate estimate. 


\section{Discussion}\label{discussion}

Our new TTM notations facilitate the formal validation of cyber-physical system requirements. In the train control system (Sec.~\ref{index:example:train}), the indexing construct allows us to select a specific actor (e.g., a train, a process, etc.) and specify a temporal property for that actor. Synchronous events, together with primed variables, allow us to check (real-time) response properties of the tabular requirements of a nuclear shutdown system (Sec.~\ref{sync:example:nop}). 

To our knowledge, the introduced notations of indexed events and synchronous events (and its combination with primed variables) are novel. For synchronous events, the conventional Communicating Sequential Processes (CSP)~\cite{Roscoe2010} and its tool~\cite{Gibson-Robinson2014} support multi-way synchronization by matching event names in parallel compositions. However, the conventional CSP does not allow processes to modify a shared state. Instead, the system state can only be managed as parameters of recursive processes, making it impossible to synchronize events that denote different parts of simultaneous updates. The notations of un-timed CSP\# and the stateful timed CSP (extended with real-time process operators such as time-out, deadline, etc.)~\cite{Sun2013} allow events to be attached with state updates. However, their semantics and tool support do not allow events that are attached with updates to be synchronized. The UPPAAL model checker and its language of timed automata~\cite{Larsen1997} support the notion of broadcast channel for synchronizing multiple state-updating transitions (one sender and multiple receivers). However, the RHS of assignments can only reference values evaluated at the pre-state. There is no mechanism, such as the notion of primed variables supported in TTM, for specifying the intended data flow. 

For indexed events, the verification tool support for both conventional CSP~\cite{Roscoe2010} and UPPAAL~\cite{Larsen1997} does not allow for fairness assumptions. For UPAAL, it is likely to manually construct an observer, but this is likely to result in convoluted encoding in larger systems and thus is prone to errors. On the other hand, the PAT tool allows users to choose fairness assumptions at the event, process, or global level~\cite{SunLDP09} for verifying the un-timed CSP\# and stateful timed CSP~\cite{Sun2013}. However, our notion of indexed events are of finer-grained for imposing fairness assumptions, as we allow the declaration of event indices as fair. 


\bibliographystyle{eptcs}
\bibliography{esss2015}

\end{document}